\documentclass{elsart}
\usepackage{epsfig}
\begin{document}

\vspace*{-.8cm}
\begin{frontmatter}
\title{Comparative Study of Multifragmentation of Gold  Nuclei
Induced by Relativistic Protons, $^4$He, and $^{12}$C}

\begin{center}
\large{S.P.~Avdeyev$^a$,  V.A.~Karnaukhov$^a$,
L.A.~Petrov$^a$\footnote{deceased},
V.K.~Rodionov$^a$, P.A.~Rukoyatkin$^a$, V.D.~Toneev$^a$,
H.~Oeschler$^b\footnote{Corresponding author: email: h.oeschler@gsi.de}$,
O.V.~Bochkarev$^c$, L.V.~Chulkov$^c$,  E.A.~Kuzmin$^c$,
A.~Budzanowski$^d$,  W.~Karcz$^d$, M.~Janicki$^d$,
E.~Norbeck$^e$, A.S.~Botvina$^f$, K.K.~Gudima$^g$}\\
{\small\it $^a$ Joint Institute for Nuclear Research, 141980 Dubna, Russia} \\
{\small\it $^b$ Institut f\"ur Kernphysik, Darmstadt University of Technology,
64289 Darmstadt,  Germany}\\
{\small\it $^c$ Kurchatov Institute, 123182, Moscow, Russia}\\
{\small\it $^d$ H.~Niewodniczanski Institute of Nuclear Physics, 31-342, Cracow,
Poland }\\
{\small\it $^e$ University of Iowa, Iowa City, IA 52242 USA  }\\
{\small\it $^f$ Institute for Nuclear Research, 117312, Moscow, Russia }\\
{\small\it $^g$ Institute of Applied Physics, Kishinev, Moldova}
\end{center}

\vspace*{-.8cm}
\begin{abstract}
Multiple emission of intermediate-mass fragments has been
studied for the collisions of p, $^4$He and $^{12}$C on Au
with the $4\pi$ setup FASA.  The mean IMF multiplicities 
(for the events with at least one IMF)
are saturating at the  value of $2.2\pm0.2$ for the  incident energies
above 6 GeV.  The observed IMF multiplicities
cannot be described in a two-stage scenario, a fast cascade
 followed by a statistical multifragmentation.
Agreement with the  measured IMF multiplicities is obtained
by introducing an intermediate phase and modifying
empirically the excitation energies and masses of the remnants.

The  angular distributions and energy spectra
from the p-induced collisions are in agreement with the scenario
of ``thermal'' multifragmentation of a hot and diluted target spectator.
In the case of $^{12}$C+Au(22.4 GeV) and
 $^4$He(14.6 GeV)+Au collisions, deviations from a pure
thermal break-up are seen in the  energy spectra of the emitted
fragments, which are harder than those both from
model calculations
and from the  measured ones for p-induced collisions. This difference
is attributed to a collective flow.
\end{abstract}

\begin{keyword}
Nuclear Reactions;
 p(2-8 GeV), $^4$He(4 and 14.6 GeV), $^{12}$C(22.4 GeV)+$^{197}$Au;
measured IMF multiplicity;
charge, energy and angular distributions; collective flow;
comparison with the model calculations.\\
PACS Numbers: 25.40-h, 25.70.Pq, 25.75-q\\

\end{keyword}
\end{frontmatter}

\section{Introduction}

Nuclear fragmentation was  discovered  in cosmic rays 60 years ago
\cite{1,2} as a puzzling phenomenon in which  nuclear fragments are
emitted from collisions of relativistic protons with various targets.
Fragments heavier than $\alpha$  particles but lighter  than
 fission fragments have been observed.
Now, they are commonly called Intermediate Mass
Fragments (IMF, $3 \leq Z \leq 20$).  Later on, in the fifties, this
phenomenon was observed in accelerator experiments \cite{3} and
 then studied leisurely for three decades. The situation  changed
dramatically in 1982 when multiple emission of IMF's was discovered
in the  $^{12}$C (1030 MeV) irradiation of emulsion at the CERN
synchrocyclotron \cite{4}. These findings stimulated the development of
many theoretical models to put forward the attractive
idea that copious production of IMF's may be related to a liquid-gas phase
transition in nuclear matter \cite{5,6,7,8,Hirsch}.
The recent status on multifragmentation can be found in Ref.~\cite{9}.

About a dozen sophisticated
experimental devices were created to investigate this process by using  heavy
ion beams, which are well suited for producing extremely hot systems. But in
the case of heavy projectiles, nuclear heating  is accompanied by
compression, fast rotation and shape distortion which may cause
dynamic effects in the multi-fragment disintegration, and
it is not easy to
disentangle all these effects and extract information on  thermodynamic
properties of hot nuclear systems.  The situation becomes
more transparent if light relativistic projectiles are
used.  In this case,  dynamic effects are expected to be negligible.  Another
advantage is that all the fragments are emitted by a single source---a slowly
moving target remainder. Its excitation energy should be
almost entirely thermal.
Light relativistic projectiles provide therefore a unique possibility for
studying ``thermal multifragmentation''.
It has been shown that thermal  multifragmentation indeed takes place
in collisions of light relativistic projectiles (p, $\bar{\rm{p}}$, $^3$He,
$^4$He, $ \pi^{-} $ ) with
a heavy target, and fragments are emitted from a diluted excited residue after
an expansion driven by thermal pressure \cite{Yen,11,12,13,Beau,Lefor}.
From IMF-IMF correlation data, fragment emission times of  less than
100~fm/c have been deduced~\cite{14,15,16}.
This value is considerably smaller than
the characteristic Coulomb time $\tau_c \approx 10^{-21}$~s~\cite{10},
 which is the mean time  for the Coulomb acceleration of
fragments.

The time scale of the IMF emission is a crucial  question
for understanding this decay mode: Is it a ``slow'' sequential
process of independent emission of IMF's or a new (multibody)
decay mode with ``simultaneous" ejection of fragments governed
by the total accessible phase space? Only the latter process
is usually called ``multifragmentation".
``Simultaneous" emission means that all fragments are liberated during
a time smaller than the characteristic one $\tau_c$.
In that case the IMF's are not emitted independently, as they
interact via Coulomb forces and are accelerated  after freeze out in a common
electric field.
Based on the measured emission times~\cite{14,15,16},
the trivial mechanism of multiple IMF emission as independent
fragment evaporation is excluded.

In this paper we present  results of the experimental study of the
multifragment emission induced by relativistic helium and carbon ions
 and compare them with our data \cite{12}
obtained for p+Au collisions. The fragment multiplicities, energy, charge
and angular distributions measured are analyzed in the framework of the
 combined approach: Cascade Model followed by the Statistical
Multifragmentation Model. Emphasis is put on the question of thermalization
and the study of a transition from a pure statistical process to a behavior
evidencing
the onset of a fast expansion as already
demonstrated in~\cite{Avd_pl}.

\section{The Experiment}

\subsection{Experimental Setup}

The experiments were performed with beams from the JINR
synchrophasotron in Dubna using the
$4 \pi$-setup FASA. Details are given in \cite{18,17}
including a brief description of the hardware and software. 
The device consists of two main parts :
First, five  $\Delta E$ (ionization chambers) $\times$
$E$ (Si)-telescopes, which serve as a trigger for the read-out of the system
allowing measurement of the charge and energy distributions of IMF's at
different angles. They are located at
$\theta = 24^\circ, \, 68^\circ, \, 87^\circ, \,  112^\circ$
and $156^\circ$ to the beam direction and together
cover a solid angle of 0.03 sr. Second, the fragment multiplicity
detector (FMD)  consisting of 64 CsI(Tl) counters
(with thicknesses around 35 mg$\cdot$cm$^{-2}$) which covers 89\% of $4 \pi$.
The FMD gives the number of IMF's in the event and their spatial distribution.

All the detectors have  been calibrated  with an alpha source
($^{241}$Am) located in the centre of vacuum chamber and 
precision-pulse generator.
The accuracy in calibrating
the energy scale is estimated to better than 5\%. This value is used as
systematic error in the fragment-energy measurements.

A self-supporting Au target 1.5 mg/cm$^2$ was located in the center
of the FASA  vacuum chamber ($\sim 1$~m in diameter).
The following beams were used: protons at energies of 2.16, 3.6 and 8.1 GeV,
$^{4}$He at energies of 4 and 14.6 GeV and $^{12}$C at 22.4 GeV.
The average beam intensity was $7 \cdot 10^8$ p/spill for protons
and helium and $1 \cdot 10^8$ p/spill for carbon
projectiles with a spill length of 300 ms and a spill period of 10 s.

\subsection{Analysis of Fragment Multiplicity}

Using the FMD array, the associated IMF multiplicity distribution $W_A
(M_A)$ is measured in events triggered by a fragment  in at least one of
the telescopes. The triggering probability is proportional to the
multiplicity $M$ of the event (primary IMF multiplicity). Hence, the
contribution of  events with higher multiplicities in $W_A (M_A)$ is
enhanced.  This is a reason why $W_A (M_A)$ differs from the
primary multiplicity distribution $W(M)$.  Another reason is the FMD
efficiency which is less than 100\%  and depends on the off-line threshold of
the scintillator counters being adjusted in such a way to reduce the
admixture of particles with $Z \leq 2$  in the counting rate of IMF's
up to the level $\leq 5$\%.  These distributions are  mutually related via
the response matrix of the FASA setup $Q(M_A$, $M)$ 
(for details see Ref.~\cite{18,17}):

\begin{equation}
W_A (M_A) = \sum \limits_{M=M_A+1}
Q(M_A, M) \cdot W(M)~~.
\end{equation}

 There are two options for obtaining the primary multiplicity distribution
$W(M)$  from the measured one $W_A (M_A)$. The first is to parameterize the
$W(M)$ distribution (using Fermi functions), to fold it with the
experimental filter according to
Eq.~(1) and then to find the parameters by the best fit to the data. 
The second option is the direct reconstruction of $W(M)$ using the inverse 
matrix  $Q^{-1}$ $(M,M_A)$. Both procedures  give rise to similar results.

\begin{figure}
\epsfig{file=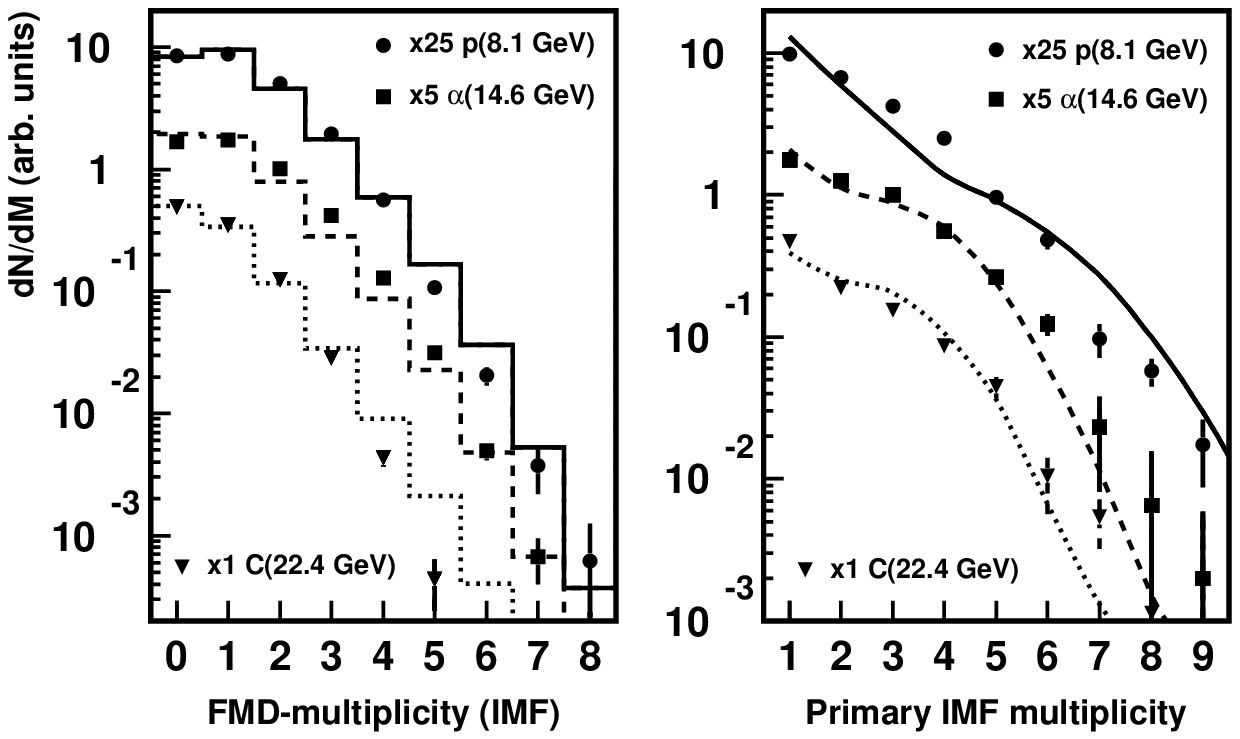,width=15cm}
\vspace*{-0.5cm}
\caption{
     a) Measured IMF-multiplicity distributions associated with
        a trigger fragment and fits with a Fermi functions (histograms) for p+Au
        collisions at 8.1 GeV (circles, solid line),
        $^{4}$He+Au at 14.6 GeV (squares, dashed line),
        and $^{12}$C + Au at 22.4 GeV (triangles, dotted line).
     b) Symbols represent
        directly reconstruct\-ed primary IMF distributions.
        The smooth lines are
        calculated with the RC+$\alpha$+SMM model.
}
\label{EXIMF}
\end{figure}

 In Fig.~\ref{EXIMF}  the multiplicity distributions obtained for   the
gold target fragmentation   by 14.6 GeV alphas and 22.4 GeV
carbon ions are compared to that for the p(8.1 GeV)+Au
collision. In these cases the mean values $<M>$ are  always about 2.1 -- 2.2
(see Table 1) being close to that obtained by the ISIS group
for $^3$He +Au collisions at 4.8 GeV \cite{19}.
Note, these values correspond to events with at least one IMF emitted.
In this definition $M$ is never less than 1. The mean multiplicity for
the all inelastic  events is smaller by the factor $[1 - P(0)]$ where
$P(0)$ is the probability to have no IMF in the collision.

\section{Model Calculations}

The reaction mechanism of multifragmentation induced by
light relativistic projectiles is usually divided
into two  steps~\cite{Bot}. The first one consists of a fast energy-deposition
stage,
during  which very energetic light particles are emitted and a nuclear
remnant (spectator) is excited. The second one is a decay of the target
spectator. The fast stage is usually described in
terms of a kinetic approach.  We use a refined version of the intranuclear
cascade model~\cite{20} to get the distributions of  nuclear remnants in
charge, mass and excitation energy.  The second stage can be described by
multifragmentation models. The Statistical Multifragmentation Model
(SMM)~\cite{21} coupled on the event-by-event basis to the cascade model is
employed here.  It will be discussed
in section 4 whether the assumption of thermal equilibrium
is justified.

\subsection{Refined Cascade Model}

The Refined Cascade Model (RC) is a version of the Quark-Gluon String
Model  developed in Ref.~\cite{23} and extended towards intermediate
energies in Ref.~\cite{24}.  This is a microscopic model which is based on
the relativistic Boltzmann-type transport equations and the string
phenomenology of hadronic interactions.  Baryons and mesons belonging to the
two lowest $SU(3)$ multiplets along with their antiparticles are included.
The interactions between the hadrons  are described by a collision term,
where the Pauli principle is imposed in the final states.  This includes
elastic collisions as well as hadron production and resonance-decay
processes.  
A fixed formation time $\tau_f = 1$ fm/c for produced particles is
incorporated. At moderate energies, in the limit $\tau_f \to 0$,
this treatment reduces to the conventional cascade model~\cite{20}.

\begin{figure}
\epsfig{file=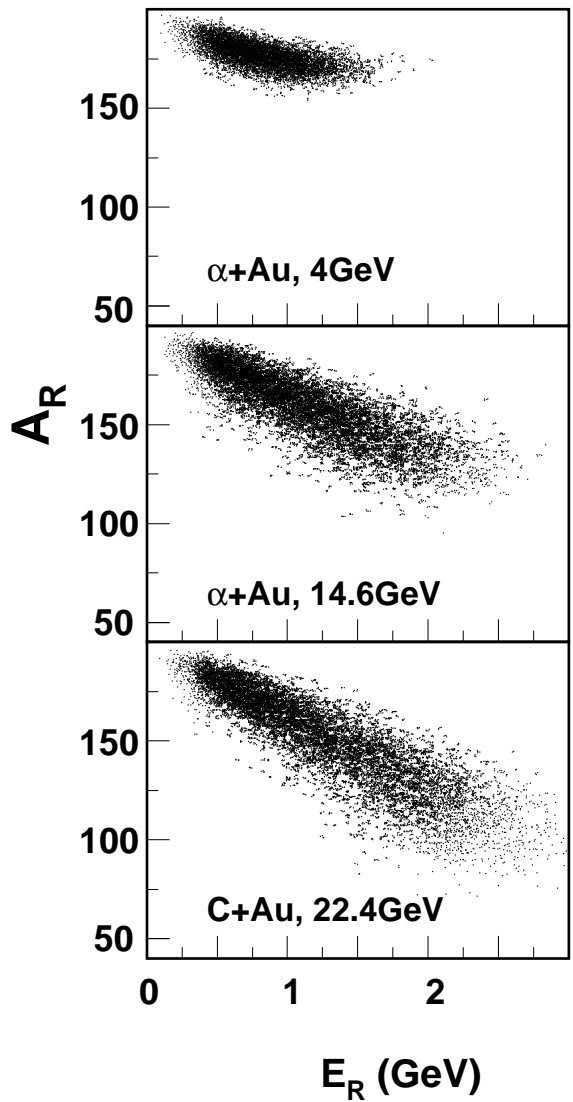,width=10cm}
\caption{
Distribution of residual masses $A_R$ and excitation energies
$E_R$ after the cascade calculation for $^4$He+Au at 4 and 14.6 GeV,
$^{12}$C+Au at 22.4 GeV.
}
\label{inc_x}
\end{figure}

Mean field dynamics is neglected in our consideration. However, we keep
the nuclear scalar potential to be defined for the initial state in the local
 Thomas-Fermi approximation, changing in time only the potential depth
according to the number of knocked-out nucleons. This ``frozen mean-field''
approximation allows us to take into account nuclear binding and the
Pauli principle as well as to estimate the excitation energy of the
residual nucleus through excited particle-hole counting.
This approximation is good for hadron-nucleus or
peripheral nucleus-nucleus collisions where there is no large disturbance
of the mean field, but it is getting  questionable for violent central
collisions of heavy ions.
Typical results for the distributions of residual masses $A_{R}$
versus their excitation energies $E_R$ in this model
are shown in Fig.~\ref{inc_x}.

It is traditionally assumed that  after completing the cascade stage the
excited residual nucleus is in an equilibrium state. In general this is not
evident. The RC includes the possibility to describe the attainment of
thermodynamic equilibrium in  terms of the pre-equilibrium (PE) exciton
model~\cite{20,25,26}. During this equilibration process some
pre-equilibrium particles may be emitted, which will result in the change of
characteristics for the thermalized residual nucleus.
The influence of this option is discussed in section 4.1.

\subsection{Statistical Multifragmentation Model}

\begin{figure}
\epsfig{file=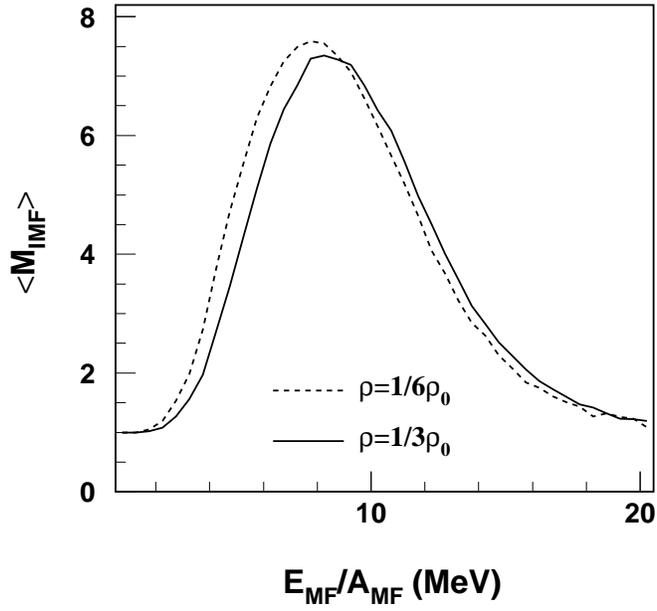,width=13.5cm}
\vspace*{-1.5cm}
\caption{
Mean IMF multiplicities (for events with at least one IMF)
as a function of the thermal excitation energy
according to SMM, calculated for freeze-out densities $\approx 1/3 \,
 \rho_0$ and $1/6 \, \rho_0$.
}
\label{INC_Bot}
\end{figure}

Within the SMM~\cite{21}, the probability of a decay
from equlibrium into the
given channel is proportional to its statistical weight.
The calculations are of Monte-Carlo type.
The break-up  volume $V_b$ is one parameter.
It is taken  as $V_b = (1 + k) A/ \rho_0$, where $A$ is the mass number of
the fragmenting  nucleus, $\rho_0$ is the normal nuclear density and
$k$ is a free parameter.
In Refs.~\cite{11,12,15} we have
shown that the break up occurs at low density.
To reach these  densities it is assumed that the system
has to expand before the break up.
Furthermore, the primary fragments may be excited
and their deexcitation is taken into account to get final IMF distributions.
Figure \ref{INC_Bot}
shows the IMF multiplicity as a function of the excitation energy
calculated for $k = 2$ and $k = 5$ which  corresponds to freeze-out
densities of $\approx 1/3 \, \rho_0$ and $1/6  \, \rho_0$, respectively.
The calculations have been performed with the RC+SMM combined model for
$^4$He+Au collisions at 14.6 GeV.  The fragment multiplicity rises with
excitation energy up to a maximum and then decreases due to vaporization of
the overheated system. This so-called ``rise and fall'' of multifragmentation
is well visible in the figure and was first demonstrated experimentally by
the ALADIN group for the collisions of $^{197}$Au at 600 MeV/nucleon with Al
and Cu targets~\cite{30}. 

Figure \ref{INC_Bot} evidences that the
choice of the break-up density only slightly influences $<M>$. The kinetic
energies of fragments are more affected because they are determined mainly
by the Coulomb field in the system depending noticeably on its size.
The use of a large value of the parameter ($k=5$) results in
the underestimation of the fragment kinetic energies compared to the
data~\cite{31}.
In the calculations presented here we used $k=2$ based on
our analysis of the correlation data~\cite{15}.

\section{Results and Discussion}

\subsection{Fragment multiplicity and excitation energy  of the system}

\begin{figure}
\epsfig{file=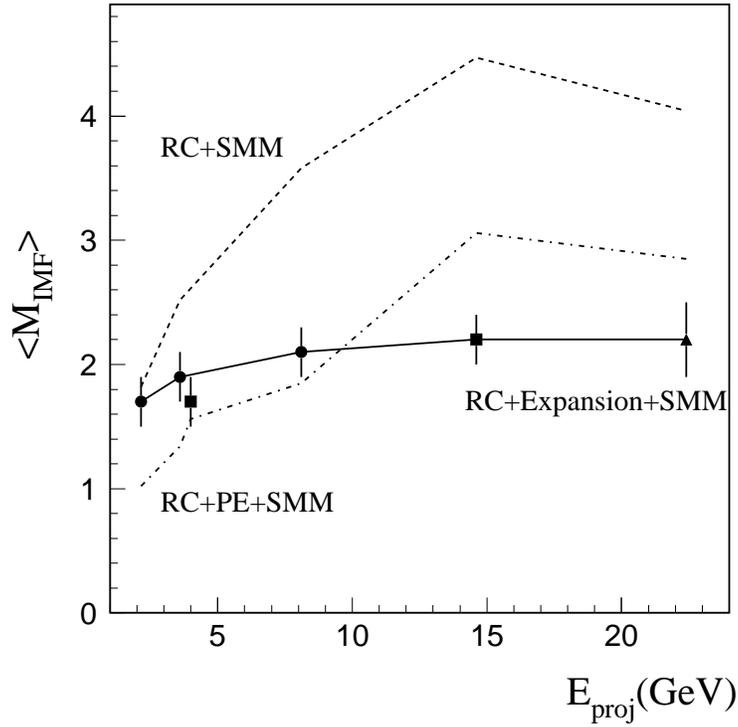,width=13.5cm}
\caption{
Mean IMF multiplicities (for events with at least one IMF)
        as a function of beam energy.
        The points are experimental data(circles - for p;
        squares - for $^4$He, triangles - for $^{12}$C. 
        Dashed and dash-dotted lines are
        drawn through the values calculated with
        RC+SMM   and with  RC+PE+SMM at the beam energies used.
        The solid line is obtained with the use of RC+$\alpha$+SMM.
        For simplicity, only one
        line is drawn for a given model calculation neglecting some
        dependence on projectile mass.
}
\label{MIMF}
\end{figure}

 The  mean IMF-multiplicities, measured and calculated, are shown
in Fig.~\ref{MIMF} as a function of the total beam energy for various
projectiles.  The data exhibit a saturation in $<M>$
for incident energies
above $\sim$ 6 \ GeV in good agreement with findings of previous
works~\cite{12,32,33}.

The dashed line in Fig.~\ref{MIMF} is obtained by means of the combined
RC+SMM model. The calculated mean multiplicities are significantly higher than
the measured ones except for the lowest beam energy.  This fact
indicates that the model overestimates the residue excitation energy.
The inclusion of pre-equilibrium  (PE) emission after the cascade stage
(RC+PE+SMM) results in a significant decrease of the excitation energy
of the fragmenting target spectator and reduces the mean IMF multiplicity
(dotted line in  Fig.~\ref{MIMF}).  However, the multiplicity reduction turns
out to be too large for $E_{proj} < 8$ GeV giving
$<M>$ much smaller than the measured ones.  One should note
that, though the calculated value of $<M>$ for the p(8.1 GeV)+Au
collisions almost coincides with the experiment, the model-predicted fragment
kinetic energies in this approach are significantly lower than the measured
ones, as  shown
in~\cite{12}.  Because the IMF energies are determined essentially by the
Coulomb field  of the source, the RC+PE+SMM model
underestimates the charge $Z$  of the target residue.
In addition, at higher $^{4}$He-beam energies, the drop in
excitation energy after the pre-equilibrium emission is not strong enough
to get the observed fragment multiplicities.  All these facts may testify
to the existence of
other mechanisms for energy loss before the IMF emission.

From the comparison of the model calculations for the fragment multiplicities
with the data we conclude that neither RC nor RC+PE are able
to describe the properties of a target spectator over a wide range
of projectiles energies.

An example of an empirical approach to this problem is given in paper
\cite{34} devoted to an analysis of the experimental data on
multifragmentation in the reactions of $^{197}$Au on C, Al, Cu and Pb
targets at $E/A$=600 MeV. The parameterized  relations (with 7 parameters)
were developed to get the
mass and energy distributions of highly excited thermalized nuclear systems
formed  as the spectator parts of the colliding nuclei. These distributions
 were used as input
for SMM calculations with the parameters adjusted to fit experimental
IMF multiplicity distributions and their yields. It should be stressed
that the suggested parameterization is specific for the considered reaction.

In our approach
we start with the results of the cascade calculation
and modify them  empirically. As discussed in \cite{35},
the excitation energies of the cascade remnants have been reduced
by factor $\alpha$ (see below) and
we assume that the drop in excitation energy is accompanied
by a mass loss. This combination holds both for preequilibrium emission,
in the spirit of the exciton model~\cite{26}, and for particle evaporation
during expansion, as considered by the EES model~\cite{22}.
The excitation
energies $E^{RC}_R$ of the residual nuclei $A_R$ given by the RC code are
reduced by a factor
$\alpha$ to get  the excitation energy of the multifragmenting state $E_{MF}$,
i.e.~$E_{MF} = \alpha \times E^{RC}_R$.  In other words, the drop in
the excitation energy is equal to $\Delta E = (1 - \alpha) E^{RC}_R$.  As
is known from the cascade calculations, $E^{RC}_R$ is proportional to the
nucleon loss during the cascade $\Delta A^{RC}$, so $\Delta E = (1 - \alpha)
\varepsilon_1 \Delta A^{RC}$, where $\varepsilon_1$ is a mean excitation
energy per ejected cascade nucleon.  The loss in mass $\Delta A$
corresponding  to this drop in excitation energy is
$\Delta A = \Delta E/ \varepsilon_2$,
where $\varepsilon_2$ is the mean energy removed by a
nucleon.  Assuming $\varepsilon_2 \approx \varepsilon_1$  one gets $\Delta A
= (1 - \alpha) \Delta A^{RC}$. We denote this empirical combined model as
RC+$\alpha$+SMM .

In the earlier paper \cite{12} for p+Au collisions, a simple
relation to determine $\alpha$ could be applied
$$
\alpha = \frac{<M_{exper}>}{<M_{INC + SMM}>} \,
$$
 because the range of the excitation energies corresponded
to  the rising
part of the energy dependence of $<M>$ shown in Fig.~\ref{INC_Bot}.
However, due to the rise-and-fall effect in $<M>$,
this relation fails for heavier projectiles. For these systems the
values of $\alpha$ are empirically adjusted to reproduce the measured
mean IMF multiplicities.
 The charge, mass and energy characteristics of fragmenting
nuclei resulting from this fitting procedure are presented in Table 1
for various colliding systems.  The corresponding values
 for the p+Au case differ slightly from those given in Ref.~\cite{12}
because a new cascade code is used here.  The values of the
parameter $\alpha$ can be obtained from Table 1 by calculating the ratio
 $E_R$(RC+$\alpha$+SMM)/$E_R$(RC+SMM)  which gives
 0.93, 0.76 and 0.53 (for p+Au),
0.49 and 0.25 (for He+Au), 0.22 (for C+Au), respectively.

\begin{table}
\begin{center}
\begin{tabular}{|c|c|c|c|c|c|c|c|c|c|c|}  \hline
$E_{inc}$& Proj  & Exper.     &\multicolumn{7}{|c|}{Calculations} & Model  \\
\cline{4-10}
(GeV) & &$M_{IMF}$&$M_{IMF}$& $Z_R$ & $A_R$ & $Z_{MF}$ & $A_{MF}$
  & $E_R$  & $E_{MF}$ &     \\
\hline
     &   &            & 1.82 & 77 & 189 & 76 & 185 & 310 & 589  & RC+SMM
\\ 2.16 & p&1.7$\pm$0.2 & 1.02 & 72 & 176 & 62 & 145 & 119 & 266   &
RC+PE+SMM    \\ &      &         & 1.69 & 77 & 188 &75  &183  & 288 & 564
  & RC+$\alpha$+SMM  \\ \hline & &               & 2.52  & 76 & 187 & 74 & 181 &
      371 & 676   & RC+SMM   \\ 3.6  & p & 1.9$\pm$0.2 & 1.34 & 70 & 171 &
55 & 134 & 148 & 385     & RC+PE+SMM   \\ &     &    & 1.89 & 75 & 184 &73
  &175  & 282 & 568 &     RC+$\alpha$+SMM   \\ \hline & &               & 3.58 &
      75 & 183 & 73 & 175 & 488 & 808   & RC+SMM   \\ 8.1 & p  & 2.1$\pm$0.2
& 1.85 & 68 & 167 & 53 & 128 & 177 & 462    & RC+PE+SMM   \\ & &  & 2.0  &
     72 & 176 & 67 & 158 & 259 & 529 &  RC+$\alpha$+SMM  \\ \hline & & & 3.89
& 75
 & 184 & 73 & 177 & 484 & 836 &   RC+SMM   \\ 4.0& $^4$He & 1.7$\pm$0.2&
1.56&68 & 167 & 54 & 130 & 176 & 428 &   RC+PE+SMM    \\ & &  & 1.77  & 73 &
 177 & 69 & 161 & 238 & 502  & RC+$\alpha$+SMM  \\ \hline & & & 4.47 & 71
& 173 &
   66 & 159 & 723 & 1132   & RC+SMM   \\ 14.6& $^4$He & 2.2$\pm$0.2& 3.06&
63& 153 & 48 & 116 & 377 & 824 & RC+PE+SMM    \\ & &  & 2.19  & 64 & 154 &
 48 & 103 & 183 & 404 &  RC+$\alpha$+SMM   \\ \hline & & & 4.04 & 67 & 163
& 64 &
153 & 924 & 1216 &   RC+SMM   \\ 22.4 &$^{12}$C&2.2$\pm$0.3&2.85 &60 & 146 &
47 & 113 & 638 &1026   & RC+PE+SMM \\ & &  & 2.17  & 59 & 139 & 41 & 86 &
207 & 415  & RC+$\alpha$+SMM   \\ \hline \end{tabular}
\caption{
 The calculated properties of  nuclear remnants from $proj$ +  Au
collisions. The  $M_{IMF}$ is the mean number of IMF's for events with at least
one IMF and  $Z_R$, $A_R$, $E_R$ are the mean charge, mass number and excitation
energy (in MeV), respectively,  averaged over all inelastic collisions, while
similar quantities $Z_{MF}$, $A_{MF}$, $E_{MF}$ are  averaged only over residues
decaying by IMF emission.
The errors in $A_{MF}$ range from 3\% to 14\% and in $E_{MF}$ from
10\% to 20\%, the lower value refers to p(8.1 GeV) and the highest to
$^{12}$C(22.4.GeV).}
\end{center}
\end{table}

\begin{figure}
\epsfig{file=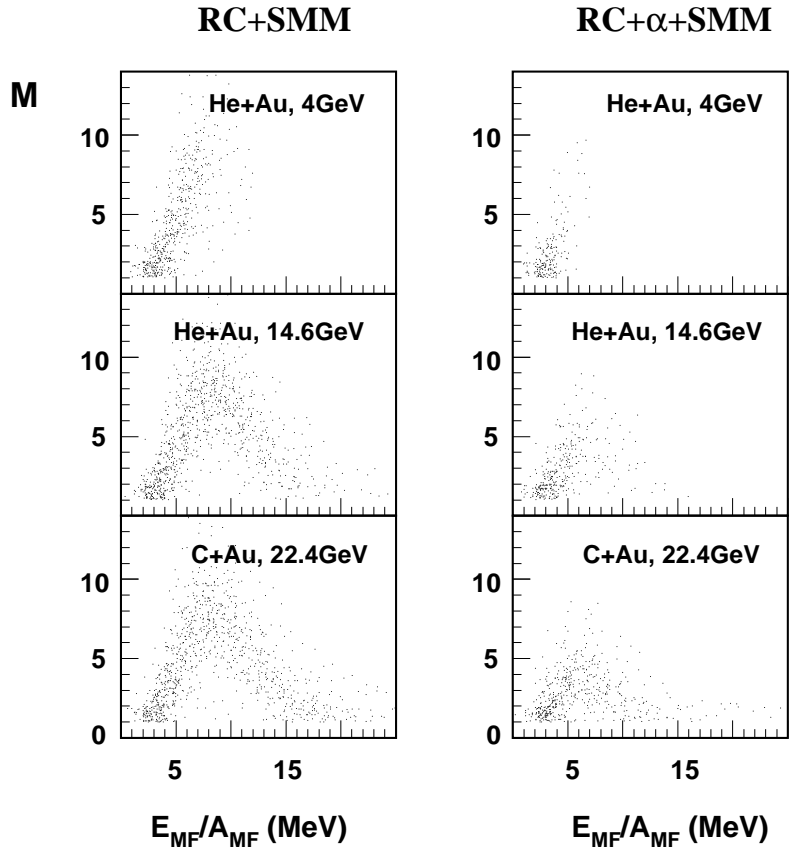,width=14cm}
\caption{
The calculated event distribution in $M$-$E_{MF}$/$A_{MF}$ plane: RC+SMM
model (left) and RC+$\alpha$+SMM approach (right).
}
\label{RC+Exp+SMM}
\end{figure}

As follows from the given values of the parameter $\alpha$,
a rather large decrease of the residual excitation energy is required
to reproduce the observed saturation effect in
$<M>$ which is caused mainly by a saturation in $E_{MF}$.  This is
illustrated in  Fig.~\ref{RC+Exp+SMM} which shows the population of events in
the $M$--versus--$E_{MF}/A_{MF}$ plane
calculated in both  the RC+SMM (left panel) and
RC+$\alpha$+SMM (right panel) scenarios.  According to the first approach
 the excitation energy distribution is rather wide and populates states
along both the rising and the falling parts of the multiplicity curve.
In the RC+$\alpha$+SMM scenario the events are mainly situated  in  the rising
part hardly approaching the region of maximal values of the IMF multiplicity.

\begin{figure}
\epsfig{file=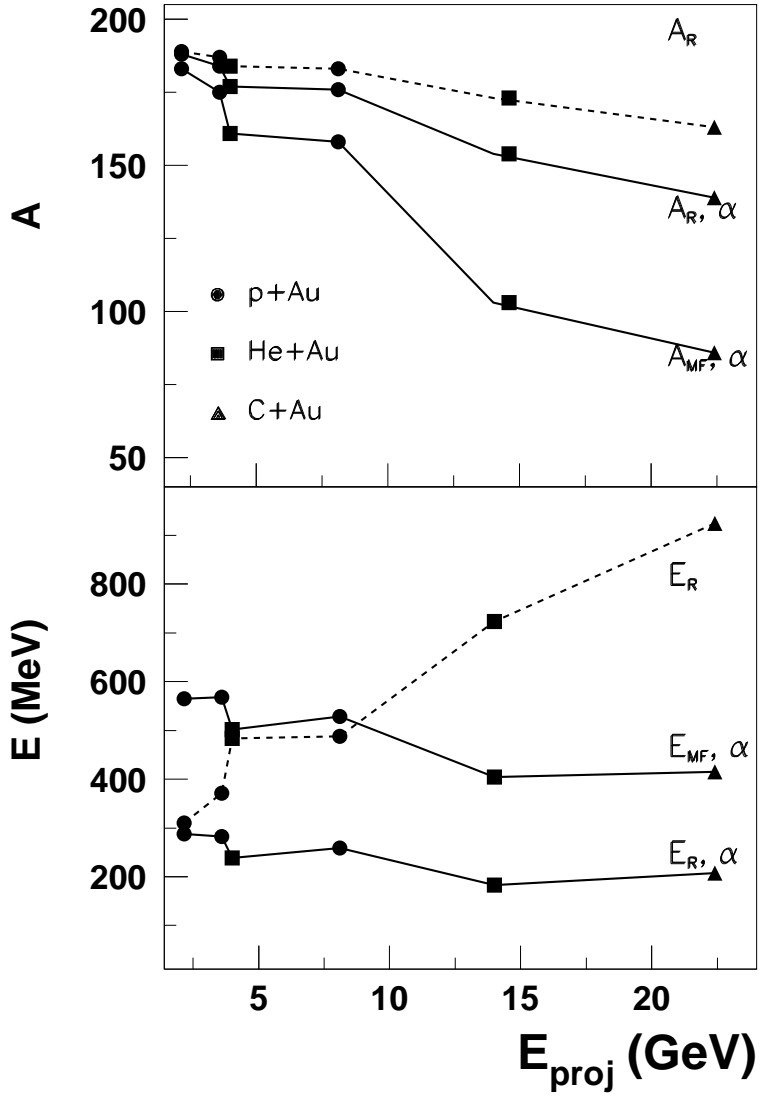,width=13cm}
\caption{
The mean values of the remnant excitation energies and
mass numbers according to Table 1: $E_R$ and $A_R$ are averaged over
all inelastic collisions, $E_{MF}$, $A_{MF}$ are for fragmenting residues.
The calculations with RC+$\alpha$+SMM are marked by ``$\alpha$".
}
\label{RC_FE}
\end{figure}

\begin{figure}
\epsfig{file=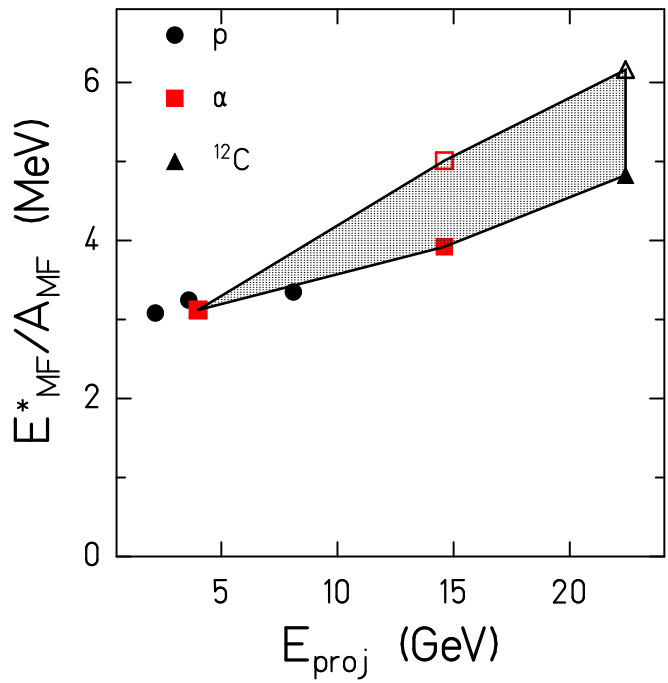,width=13cm}
\caption{Mean excitation energy of the fragmenting nucleus per nucleon
$E^*_{MF}/A_{MF}$ as a function of the beam energy:
the solid points refer to the thermal part,
in some cases the flow energy added is shown as open symbols
and grey area.}
\label{onset_of_flow}
\end{figure}

The calculated mean residual excitation energies and mean mass
numbers are presented in Fig.~\ref{RC_FE}.
The total excitation energy of the fragmenting
nucleus  $E_{MF}$ changes slightly with an increase of the incident energy.
At the same time, the excitation energy per nucleon goes up while the
residual mass
decreases, keeping the mean IMF multiplicity almost constant.  Note that the
energies given in Fig.~\ref{RC_FE} and Table 1 are thermal ones by definition.
A possible non-thermal contribution will be discussed at the end of this
section along with Fig.~\ref{onset_of_flow}.

It is of interest to compare the extracted masses and excitation
energies of fragmenting nuclei to those obtained by the EOS
collaboration for Au(1 GeV/nucleon)+C collisions
(in inverse kinematics) \cite{36}. In that paper the mass and energy
balance relations were applied using of the measured kinetic energies
of all outgoing charged particles after separating from the prompt stage
of the reaction. The neutron contribution was taken into
account on the basis of cascade and statistical model simulations.
The inclusive data were not presented there and only  the values, corresponding
to our mean IMF multiplicity were used for the comparison.
Our value of $E_{MF}$/$A_{MF}$
is close to that from Ref.~\cite{36} if a collective energy
is added (see later). As to the mean mass $A_{MF}$, the value obtained in the
present work ($\approx 90 $) is remarkably lower, because of the larger
mass loss induced by a projectile with twice the energy.

\begin{figure}
\epsfig{file=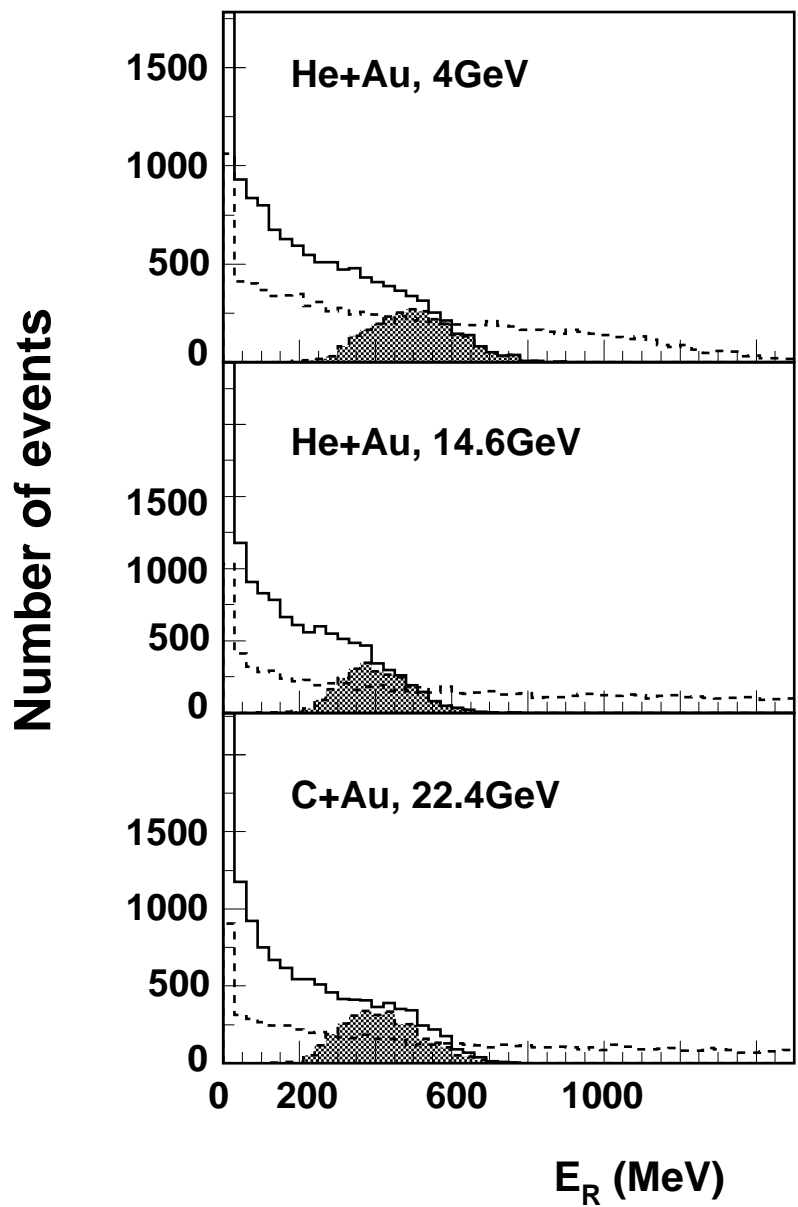,width=13cm}
\caption{
Distribution of excitation energies obtained with
RC (dashed lines), with RC+$\alpha$ (solid lines) and
the fraction decaying by multifragmentation according to RC+$\alpha$+SMM
(hatched area).
}
\label{Ex_dist}
\end{figure}

 Some examples of the excitation energy distributions are displayed
in Fig.~\ref{Ex_dist}.
The IMF emission takes place on the tail of the distributions 
(hatched area), therefore
the mean excitation of the fragmenting nuclei is
much larger than that averaged over all the target spectators.

\subsection{Angular distributions}

\begin{figure}
\epsfig{file=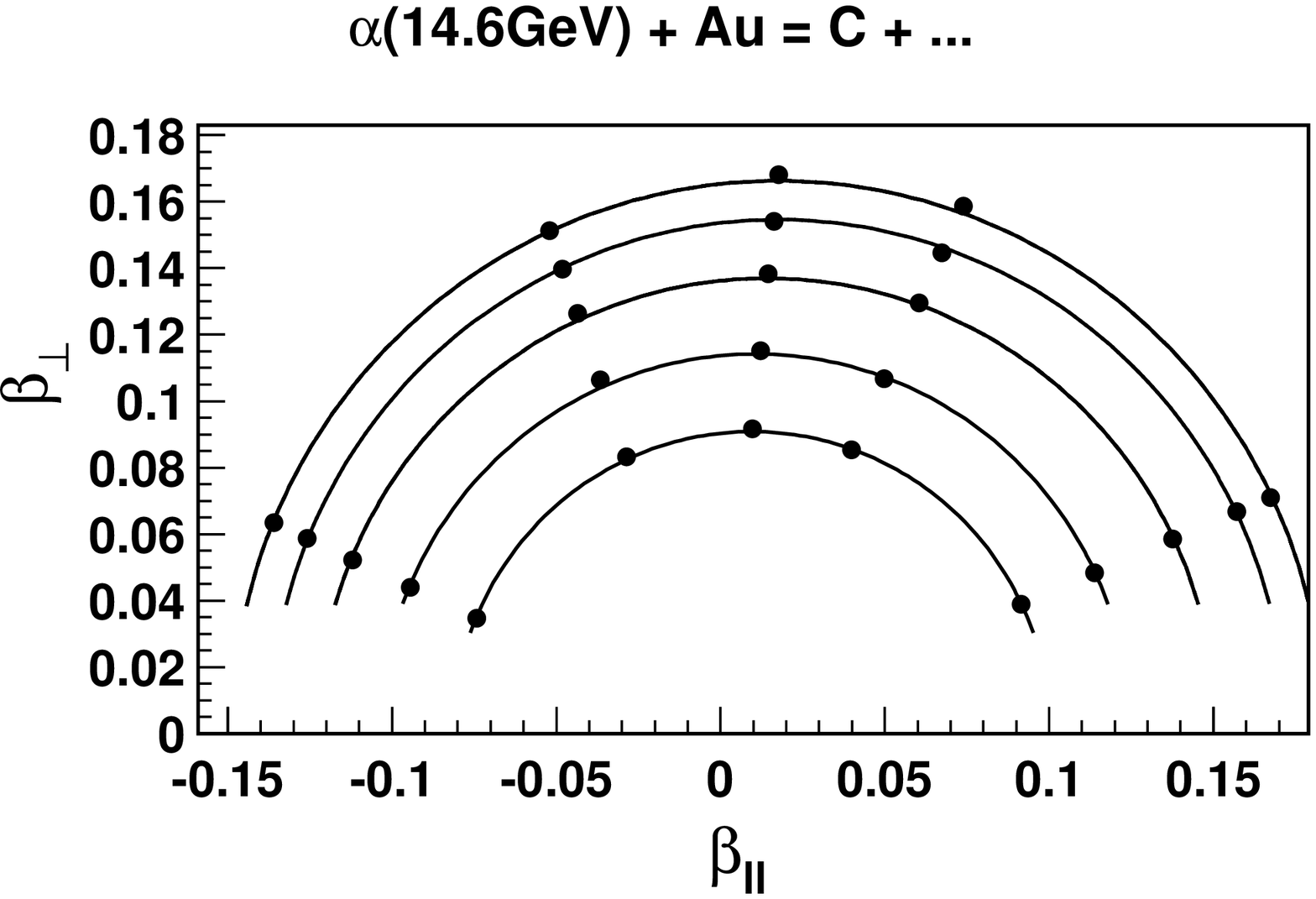,width=6.5cm}
\epsfig{file=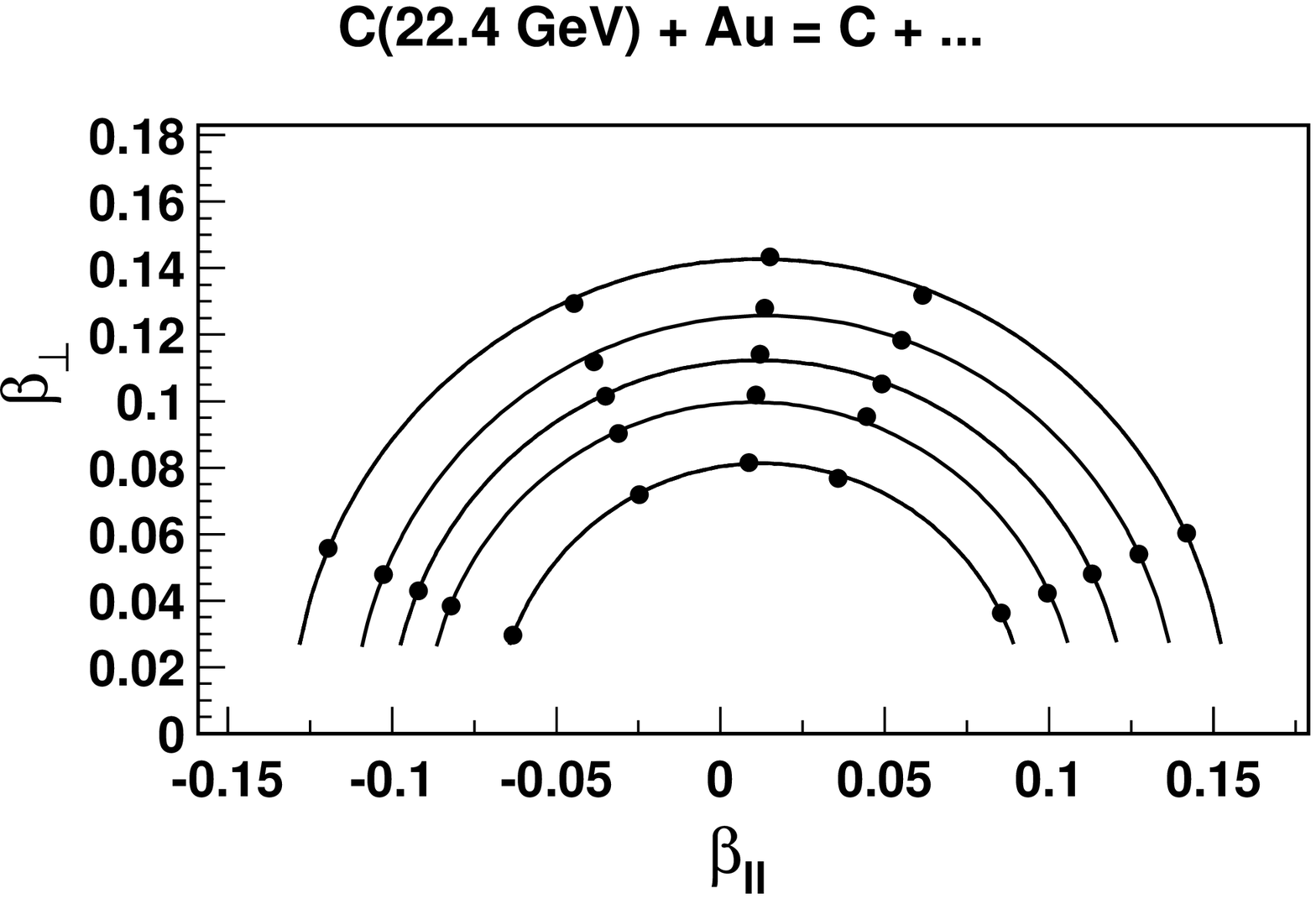,width=6.5cm}
\caption{
Transverse versus longitudinal velocity plot for emitted
carbon isotopes
produced in $^4$He(14.6 GeV) and
$^{12}$C(22.4 GeV) collisions with a Au target.
Circles are drawn through points of equal invariant cross
section corresponding to isotropic emission of the fragments in the
moving source frame.
}
\label{velocity}
\end{figure}

Now let us consider the thermalization of the system at
break-up.  To check whether
this state of the emitting system is close to
thermal equilibrium, the plot of the fragment invariant probability
distribution
in terms of the longitudinal--versus--transversal
velocity components is  presented
 in Fig.~\ref{velocity} for the $^4$He+Au and C+Au collisions.
Circles connect experimental points of equal invariant cross sections
for emitted carbon fragments in the energy range above the spectral peak.
This demonstrates an isotropic emission in the frame of a moving source
and indicates that the fragment emission proceeds
from a thermalized state with the center positions of the circles
determining the longitudinal source velocity, $\beta_{source}$.
Their mean values are close to estimates with the
RC+$\alpha$+SMM model for all cases
except $^4$He+Au at 4 GeV, where the calculations underestimate the source
velocity by almost a factor of two.
The calculated mean $\beta_{source}$ are equal
to $0.76 \cdot 10^{-2}$,
$1 \cdot 10^{-2}$, $1.36 \cdot 10^{-2}$ and $1.7 \cdot 10^{-2}$ for
p(8.1 GeV)+Au, $^4$He(4 GeV), $^4$He(14.6 GeV)+Au and C(22.4 GeV)+Au
collisions, respectively.

\begin{figure}
\epsfig{file=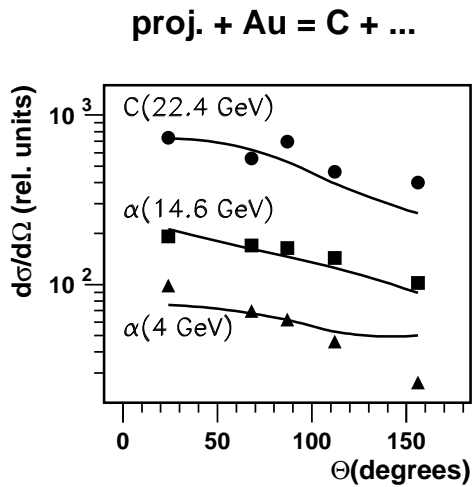,width=13cm}
\vspace*{-4cm}
\caption{
Angular distributions of carbon (in laboratory system) for $^4$He+Au and
$^{12}$C+Au collisions. The lines are calculated with RC+$\alpha$+SMM.
}
\label{angular}
\end{figure}

The fragment angular distribution in the laboratory system exhibits a
forward peak caused by the source motion as exemplified  in
Fig.~\ref{angular} for carbon fragments.
The data are well reproduced by the model calculations except
those for the helium beam  at the lowest energy. The measured distribution
here is more forward peaked, which may be considered as an indication that
the momentum transfer is larger than predicted. This observation may indicate
a stronger stopping than predicted for this case.

\subsection{Charge distributions}

\begin{figure}
\epsfig{file=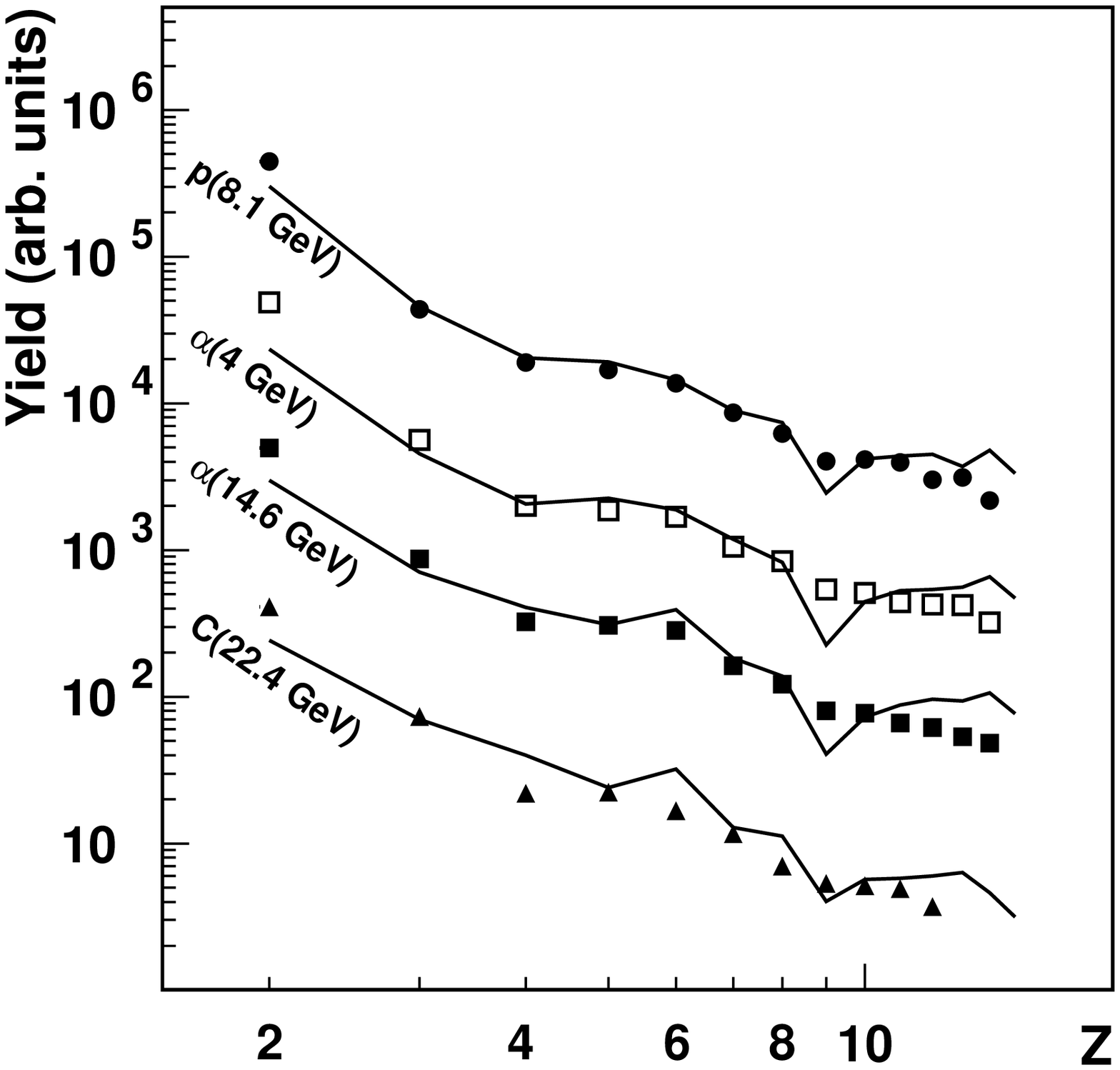,width=7.5cm}
\epsfig{file=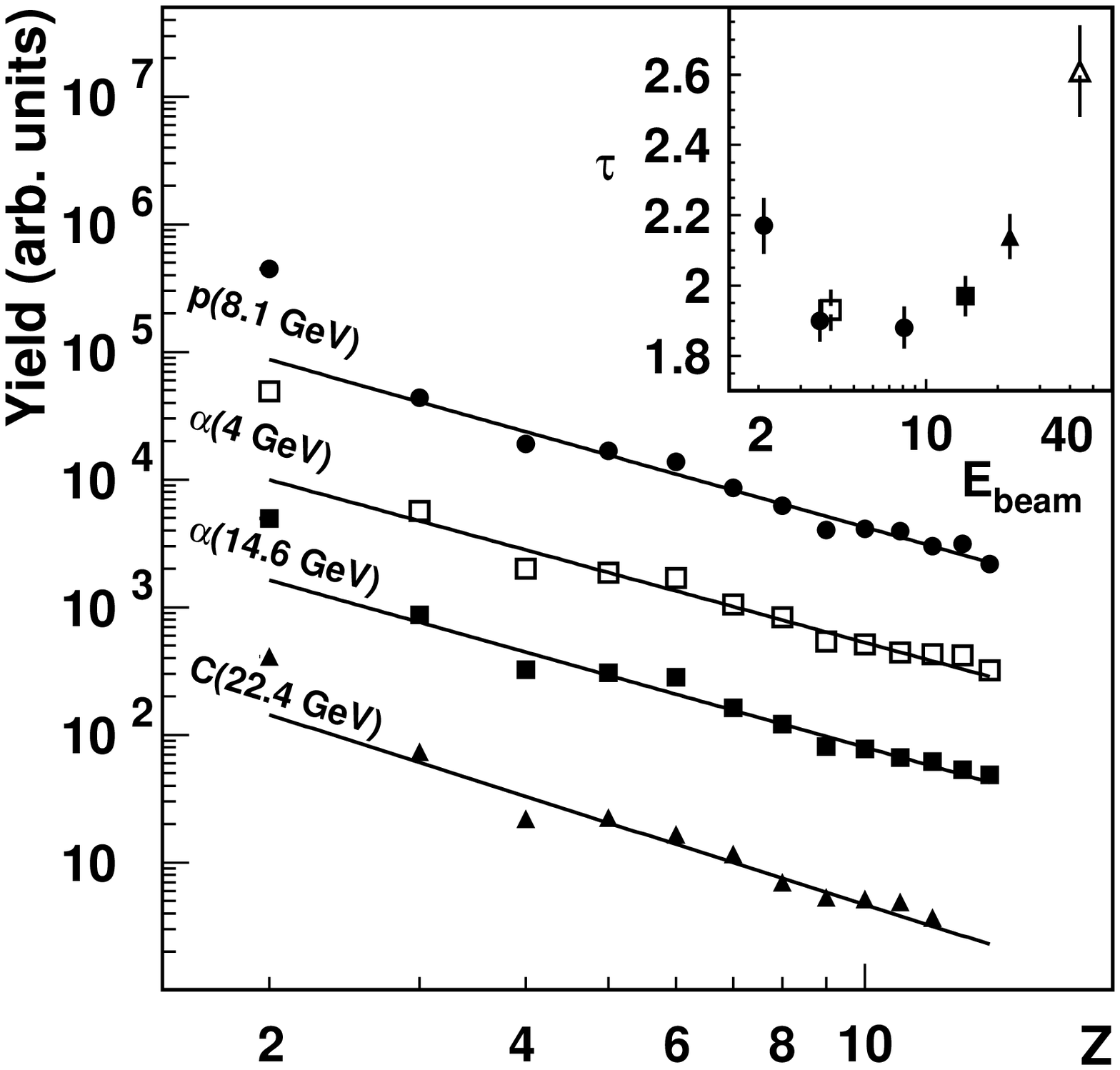,width=7.5cm}

\caption{
Fragment charge  distributions  obtained at $\theta = 87^\circ$ for
          p+Au at 8.1 GeV (top),
         $^4$He+Au at 4 GeV,
         $^4$He+Au at 14.6 GeV and
         $^{12}$C+Au at 22.4 GeV.
        The lines (left side) are calculated by \ RC+$\alpha$+SMM
        (normalized at $Z$=3). The power law fits are shown on the right panel
        with $\tau$-parameters given in the insert as a function of beam
        energy.
        The last point in the insert is for $^{12}$C+Au collisions at 44 GeV
        (from a preliminary experiment).
}
\label{charge}
\end{figure}

The  charge distributions of IMF's are shown in Fig.~\ref{charge}.
The calculations for
the RC+$\alpha$+SMM scenario agree nicely with the data
(left side of Fig.~\ref{charge}).
The general
trend of the IMF
charge (or mass) distributions is well  described by a power law $Y(Z) \sim
Z^{- \tau}$. The obtained values of the  exponent are
$\tau = 2.17 \pm 0.08$, $1.90 \pm 0.06$ and $1.88 \pm 0.06$
for proton beam of 2.1 GeV, 3.6 GeV and 8.1 GeV,
$\tau = 1.93 \pm 0.06$ and $2.0 \pm 0.1$  for helium beam of 4
GeV and 14.6 GeV and finally $\tau = 2.1 \pm 0.1$ for
carbon projectiles (Fig.~\ref{charge},
right panel).

In earlier papers on multifragmentation \cite{5,Hirsch}
this power-law behavior of  the IMF yield was interpreted  as an
indication of the proximity of the decaying state to the critical
point for the liquid-gas phase transition in nuclear matter.
This was stimulated by the application of the classical Fisher's droplet
model~\cite{39}, which predicted a pure power-law droplet-size
distribution with $\tau$=2-3 at the critical point.
According to Ref.~\cite{40} the fragmenting system is not close
to the critical point. 

The power law is well explained at temperatures far below
the critical point. As seen in Fig.~\ref{charge},
the pure thermodynamical SMM
predicts  that the IMF  charge distribution is
very close to a
power law at freeze-out temperatures of 5--7 MeV, while the critical
temperature $T_c$ (at which the surface tension vanishes) is assumed 
to be 18~MeV\null. In Ref.~\cite{41}, it was also shown that several 
results concerning the fragment-size distribution (a power-law behaviour) 
can be rendered well by the use of the kinetic
model of condensation beyond the vicinity of the liquid-gas critical point. 
The critical temperature and density of nuclei were calculated in
Ref.~\cite{6} using a Skyrme effective interaction and finite-temperature
Hartree-Fock theory. The values of $T_c$ were found to be in the range 
(8.1-20.5)~MeV
depending on the Skyrme-interaction parameters chosen.

Thermal multifragmentation is consistent with a
first-order phase transition of nuclear matter 
characterized by the liquid-gas type phase instability.
Indeed, it is consistent with experiments that fragmentation takes place
after expansion driven by thermal pressure \cite{11,12,13},
and that the decomposition time
is short (less than 100 fm/c)~\cite{14,15,16}.
In fact, the final state of this transition looks like a nuclear
fog~\cite{38}~: the liquid drops of IMF's surrounded by a gas
of nucleons and light clusters,  $d, t$ and $\alpha$-particles.
This interpretation is in the line of the
SMM~\cite{42}. Recently, several theoretical and experimental
papers have been published on that topic (see for example \cite{spinodal}).
An interesting aspect of this concept is the
isospin fractionation as was demonstrated in Ref.~\cite{isospin_fractionation}.

\subsection{Energy spectra of fragments}

In general, the kinetic energy of fragments is determined by the sum of
four terms: thermal motion, Coulomb repulsion, rotation and  collective
expansion energies  of the system at freeze out:
$E = E_{th} + E_C +E_{rot} + E_{flow}$.  The additivity of the first three
terms is quite obvious.  For the last term, its independence  of the others
may be considered only approximately when the evolution
of the system  after the freeze-out point is driven only by the Coulomb force.
The Coulomb term is significantly larger than the thermal one. It was shown in
Ref.~\cite{15} that the Coulomb part of
the mean energy of the carbon fragment is three times
larger than thermal energy.  These calculations were performed within the
RC+SMM scenario with the volume emission of fragments from a diluted system.

\begin{figure}
\epsfig{file=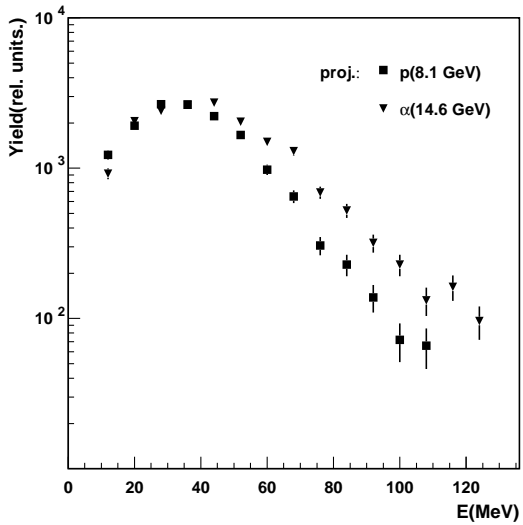,width=8cm}
\caption{Comparison of energy spectra of carbon fragments from
p+Au at 8.1 GeV and $\alpha$+Au at 14.6 GeV.}
\label{p_alpha}
\end{figure}

\begin{figure}
\epsfig{file=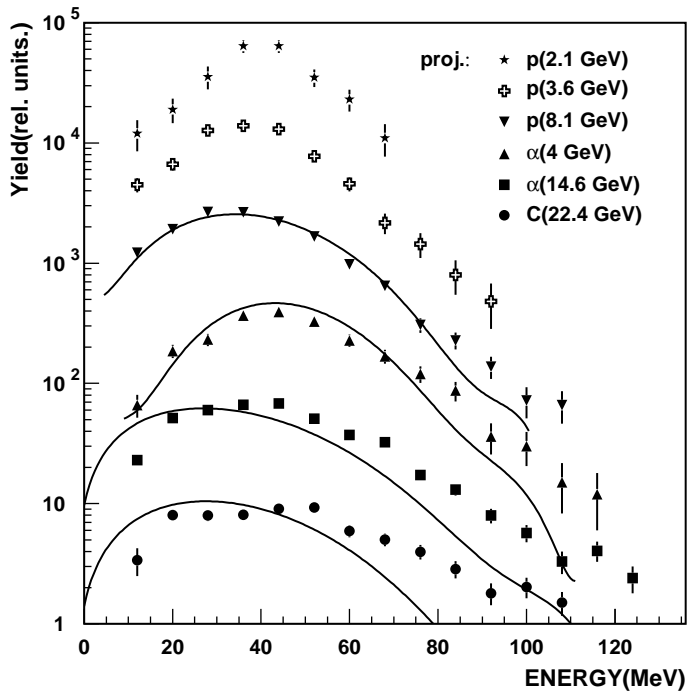,width=10cm}
\caption{
Energy distribution of carbon fragments obtained  for
different collision systems at $\theta = 87^\circ$.
The lines are calculated in RC+$\alpha$+SMM model assuming no flow.
}
\label{all_spectra}
\end{figure}

The contribution of the collective flow for the p+Au collisions
at 8.1 GeV incident energy was estimated in Ref.~\cite{12}.
This was done by comparing  the measured IMF spectra with the  calculated
ones in the framework of the SMM which  includes no  flow.
This analysis did not reveal any significant
enhancement in the measured  energy spectra  restricting   the mean
flow velocity $v_{flow}$ to less than  0.02~$c$.
For the case of heavy ion collisions,
collective flow has been observed and it is the most pronounced in
central Au+Au collisions \cite{45}.
In this respect it would be quite interesting to analyse the fragment
spectra from  He+Au and C+Au collisions looking for a possible
manifestation of collective flow.
Indeed, a comparison of the energy spectra of carbon fragments emitted in
p(8.1 GeV)+Au and $\alpha$(14.6 GeV)+Au
shows a drastic difference as demonstrated in Fig.~\ref{p_alpha}.
A more general comparison of the carbon
spectra for  proton-, helium- and carbon-induced collisions on the Au target
are presented in Fig.~\ref{all_spectra}.
The calculated carbon spectrum for p+Au collisions (at 8.1 GeV) is consistent
with the measured one. A similar situation occurs with $^4$He+Au collisions
at 4 GeV, but not with $^4$He(14.6 GeV)+Au and $^{12}$C(22.4 GeV)+Au 
interactions: the measured spectra are harder than the calculated ones.

\begin{figure}
\epsfig{file=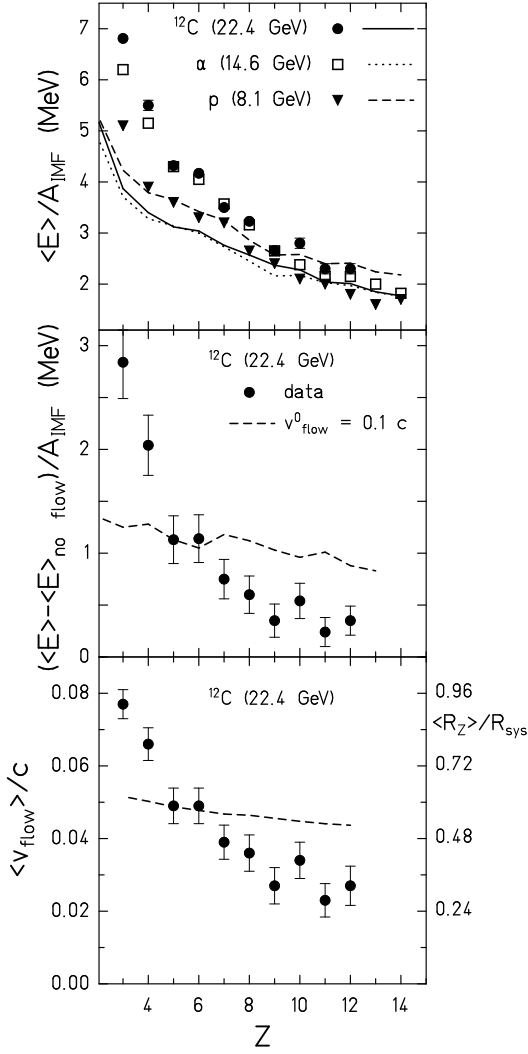,width=8cm}
\caption{Upper part:
Mean kinetic energies of outgoing fragments per nucleon
measured at $\theta = 87^\circ$ for
p(8.1 GeV), $^4$He(14.6 GeV) and $^{12}$C(22.4 GeV) collisions with Au.
The lines are calculated within RC+$\alpha$+SMM approach assuming no flow.
Middle part: Flow energy per nucleon (dots) obtained as a difference
of the measured
fragment kinetic energies and the values calculated under assumption
of  no flow in the system. The dashed line represents
a calculation assuming a linear radial profile for the
expansion velocity
with  $v^0_{flow} = 0.1$~c.
Lower part: Experimentally deduced mean flow velocities (dots)
for $^{12}$C+Au collisions  as a function of the fragment charge (left scale),
and the mean relative radial coordinates of fragments (right scale), obtained
under assumption of a linear radial profile for the expansion velocity.
The dashed line shows the mean radial coordinate according to SMM.}
\label{E_mean}
\end{figure}

The measured mean kinetic energies per nucleon $<E>/A_{IMF}$
are given in the upper part of Fig.~\ref{E_mean},
only statistical errors are shown. Systematic errors ($\sim$5$\%$) are 
the same for all the cases. They can be neglected when comparing data 
obtained for different collision systems.
There is a remarkable enhancement  in the reduced kinetic energy for
light fragments from He(14.6 GeV)+Au and C(22.4 GeV)+Au collisions as 
compared to the
p(8.1 GeV)+Au case. The calculated values of the mean fragment energies
(shown by lines) are obtained with the RC+$\alpha$+SMM model by
multibody Coulomb trajectory calculations on an event-by-event basis.
In the initial state  all  charged particles are assumed to have
a thermal velocity only. The measured energies are close to the calculated
ones for p+Au collisions in the range of the fragment charges between 4 and 9.
However, for the  $^4$He+Au and $^{12}$C+Au interactions the experimental
data exceed the calculated values.

The observed deviation is not caused by any methodical distortion of the
spectral shape for heavier beams, e.g. by the pile-up effect. Its magnitude is
determined by the counting rates, which are comparable for p and He beams and
are smaller for C + Au collisions.

This deviation cannot be attributed to an angular momentum effect.
This has been estimated in Ref. \cite{Avd_pl}
to be $<E_{rot}> / A_{IMF} \approx 0.04$~MeV/nucleon,
which is an order
of magnitude smaller than the energy enhancement for light fragments.
We suggest that this enhancement is caused by the expansion
of the system, which is assumed to be radial as the velocity plot
(Fig.~\ref{velocity})
does not show any significant deviation from circular symmetry.

\begin{figure}
\epsfig{file=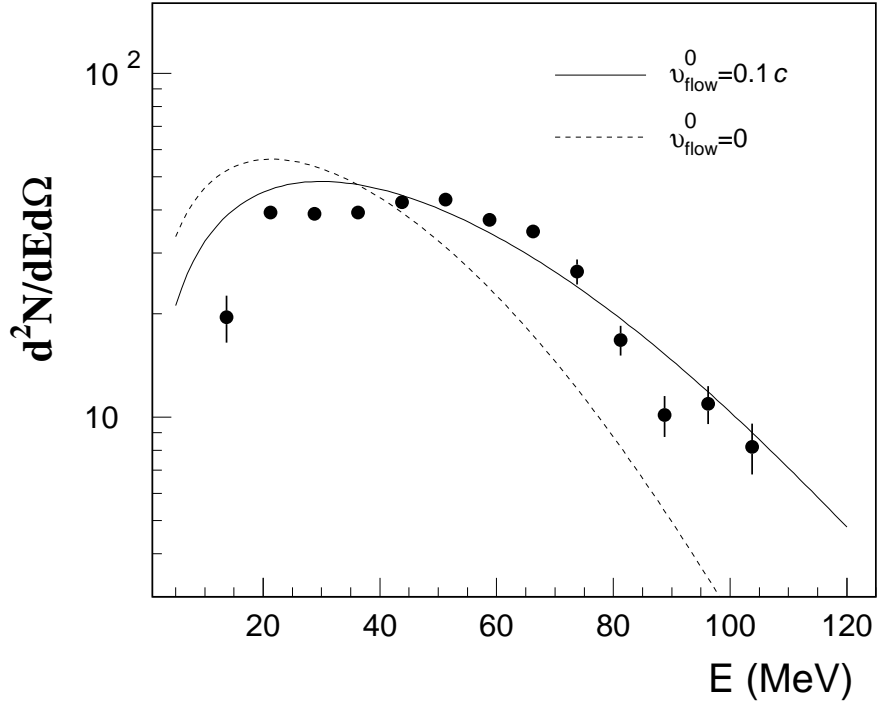,width=13cm}
\caption{Energy distribution of carbon fragments (at $\theta = 87^\circ$) from
$^{12}$C+Au collisions. Solid line is calculated assuming the radial flow with
velocity on the surface equal to 0.1$c$. Dashed line is calculated assuming no
flow.}
\label{sp_flow}
\end{figure}

An estimate of the fragment flow energy may be obtained as the difference
between the measured IMF energies and those calculated without taking into
account any flow in the system. This difference for C+Au collisions is shown
in the middle part of Fig.~\ref{E_mean}.
The error bars include both statistical and systematic
contributions.
In an attempt to describe the data we modified the SMM code in the
RC+$\alpha$+SMM concept by the inclusion of a radial velocity boost for
each  particle at freeze out. In other words, a radial expansion
velocity was superimposed on the thermal motion in the calculation
of the multibody Coulomb trajectories.
A self-similar radial expansion is assumed, where the local flow
velocity depends linearly on the distance of the particle
from the centre of mass. The expansion velocity of a particle with charge
$Z$ located at radius $R_Z$ is given by
\begin{equation}
\vec{v}_{flow} (Z) = v_{flow}^0 \cdot \frac{\vec{R}_Z}{R_{sys}}
\end{equation}
where $v^0_{flow}$ is the radial velocity at the
surface of the system. Note, that in this case the density distribution
is changing in a dynamic evolution by a self-similar way being a function
of the scaled radius $R_Z/R_{sys}$.
The use of the linear profile for the radial velocity is motivated
by hydrodynamic model calculations of an expanding hot nuclear system
(see for example Ref.~\cite{44}).
The value of $v^0_{flow}$ has been adjusted to describe
the mean kinetic energy measured for carbon fragments. 
Figure \ref{sp_flow} shows the
comparison of the measured  and calculated energy spectra of carbon fragments
assuming $v^0_{flow}$=0.1$c$. The agreement is very good.
The calculation without flow deviate strongly.

There is a longstanding problem of the qualitative 
difference between the chemical or thermal equilibrium temperatures and the 
kinetic or, so called, slope temperatures. The last discussion on that point 
can be found in \cite{Odeh}. The mean equilibrium temperature obtained in 
our calculations is 6.9 MeV. At the same time, the slope temperature 
(inverse slope parameter) extracted 
from the spectra for pure thermal decay is $T_s$=14.5 MeV 
(dashed curve). That is the mutual result of the thermal
motion, Coulomb repulsion during the volume disintegration and the secondary 
decay of the excited fragments. Introducing rather modest radial flow (with
$v^0_{flow}$=0.1$c$) results in increase of the slope temperature up to 
$T_s$=24 MeV. 

Lets return to Fig.~\ref{E_mean} middle. Dashed line presents the difference of
calculated fragment energies obtained for $v^0_{flow}$=0.1$c$ and 
$v^0_{flow}$=0.

The data deviate significantly from the calculated values for Li and Be.
This may be caused in part by the contribution of particle emission
during the early stage of expansion from the hotter and denser system.
It is supported by the fact that the extra energy of Li fragments
with respect to the calculated value is clearly seen in Fig.~\ref{E_mean}
even for proton-induced fragmentation, where no
significant flow is expected. This peculiarity of  light fragments
has been noted already by the ISIS group for   $^3$He+Au   collisions
at 4.8 GeV \cite{19}.

For fragments heavier than carbon, the calculated curve in
the middle part of Fig.~\ref{E_mean}
is above the data and decreases only slightly with
increasing fragment charge. Such a behaviour  is expected.
The mean fragment flow energy
is proportional to $<R_Z^2> $. This value is only slightly changing with
fragment charge in the SMM code due to the assumed equal probability for
fragments of a given charge  to be formed at any point in
 the available break-up volume. This assumption is a consequense of the
simplification of the model to consider the system as uniform with
$\rho(r)$=const for $r\leq R_{sys}$. The discrepancy between the data and the
calculations in Fig.~\ref{E_mean} indicates that a uniform density 
distribution is not fulfilled.
The dense interior of the expanded nucleus is
favored for the appearance of larger IMF's, if
fragments are formed via the density fluctuations.
This observation is in accordance also with
 the analysis of the mean IMF energies performed in Ref.~\cite{Hirsch,12}
for proton-induced fragmentation. It is seen also in Fig.~\ref{E_mean},
that
for p+Au collisions the measured energies are below the theoretical
curve for fragments heavier than Ne. This may be explained by the
preferential location of the heavier fragments in the interior region
of the freeze out volume, where the Coulomb field is reduced.
The deviation of data from the calculations becomes less, but still
remains, if one assumes a quadratic radial profile for the expansion
velocity.

For the estimation of the mean flow velocities of fragments, the difference
between the measured IMF energies and calculated ones (no flow)
has been used as discussed already in Ref.~\cite{Avd_pl}.
The results are presented in the lower part of Fig.~\ref{E_mean}.
The values for Li and Be are considered as upper limits because
of the possible contribution of preequilibrium emission.
    The corresponding values of
$<R_Z> / R_{sys}$, obtained under the assumption of
a linear radial profile for the expansion velocity, can be read on the
right-hand scale of the figure.
Again the mean radial coordinate for the carbon fragment is chosen to coincide
with the calculated one. The dashed line shows
the mean radial coordinates of fragments according to the SMM code. As it has
been noted above, the calculated values of $<R_Z> / R_{sys}$ are only slightly
decreasing with $Z$ in contrast with the data.

The flow velocity on the system surface obtained in this analysis is close to
the value of the maximum expansion velocity (0.08c) calculated with EES for
$^{12}$C+Au collisions \cite{Friedman_2000}.

Effects of the  radial collective energy for
1 GeV/nucleon Au+C collisions
(in inverse kinematics) were considered in~\cite{46}
by analyzing  the transverse kinetic energies $E_t$ of fragments with
$Z = 2-7$.
This was done for two charged-particle multiplicity bins,
corresponding to peripheral (M1) and central (M3) collisions.
The Berlin statistical model code~\cite{47}
was used with inclusion of a radial velocity chosen properly to account
for the experimental values of $<E_t>$.
In the case of peripheral collisions the  obtained expansion velocities
are close to those extracted here,
but the corresponding
mean IMF multiplicities (in our definition) are lower than 1.5.
For central collisions ($<M_{IMF}> \simeq 4$) the expansion
velocities are $\sim 1.5$   times higher.
It would be desirable to compare our data with those
for the intermediate case (bin M2), which are unfortunately not available.
Making an interpolation, one may see that our analysis gives slightly
lower values of $v_{flow} (Z)$ as compared to Ref.~\cite{46}.
This  may be caused by the fact that the Berlin statistical model
underestimates the Coulomb
part of the fragment kinetic energy
(see~\cite{31}) as the freeze-out density used is much smaller
($\rho_f = 1/6 \rho_0$).

The total expansion energy can be estimated by integrating the nucleon flow
energy (taken according to  Eq.~(2)) over the
available volume at freeze out. For a uniform system one gets:
\begin{equation}
E_{flow}^{tot} =
{3 \over 10} A \cdot m_N \ (v^0_{flow})^2 \ (1- r_{N}/R_{sys})^5
\end{equation}

\noindent where $m_N$, $r_N$ are the nucleon mass and radius.
For $^{12}$C+Au collisions it gives $E_{flow}^{tot} \simeq 115$~MeV,
corresponding to the flow  velocity on the surface equal $ 0.1c$.
Similar results are obtained for $^{4}$He(14.6 GeV)+Au collisions.

The excitation energy of the fragmenting systems consists of a thermal part
$E_{MF}$ given in Table 1 and, in some cases, an additional dynamical part
$E_{flow}^{tot}$. These findings are reflected in Fig.~\ref{onset_of_flow}
where the total excitation energy $E_{MF}^*/A_{MF}$ is shown
as a function of the incident energy.
The full symbols correspond
to the thermal part of the excitation energies. These values exhibit a rather
weak rise with increasing incident energy.
Adding the flow energy (grey area, open symbols) yields a stronger increase.
The onset of the collective flow takes place at excitation energies around
4 MeV/nucleon, which is in good agreement with the results of \cite{Beau}.
In contrast to Ref.~\cite{Lefor}, we do not find arguments that this onset
should be related to a transition from a surface to a bulk emission.

\section{Conclusions}

The emission of intermediate mass fragments has been studied
in the reactions p(2.1, 3.6 and 8.1 GeV)+Au, $^4$He(4 and 14.6 GeV)+Au
and $^{12}$C(22.4 GeV)+Au.
The measured IMF multiplicities (for events with at least one IMF)
saturate at a value of $2.2\pm0.2$ for incident energies above 6 GeV.
This saturation in IMF multiplicity
seems to correspond to total excitation energies around 3 -- 4 MeV/nucleon.
The angular distributions of the IMF's are slightly forward peaked;
the yield distributions of parallel versus perpendicular
velocities exhibit circular symmetry. These results show
that the IMF's are emitted from a source that moves with
rather low velocity (0.01 -- 0.02) $c$
and support the interpretation of ``thermal
multifragmentation'', a break-up of a diluted system.

Model calculations for the IMF multiplicities using a two-stage concept,
a cascade followed by SMM (statistical multifragmentation model),
fail to describe the measured values.
Taking into account pre-equilibrium particle emission before
attainment of thermal equilibrium in the system decreases the number of
IMF's, but this is still not sufficient for describing the observed
multiplicity saturation. 
Only if one applies
an empirical modification of the calculated excitation energies $E_R$ and
residual masses $A_R$ after the cascade used as input for the SMM
calculations, the observed saturation of the IMF multiplicity can be
reproduced.  This study shows  that an intermediate step is needed which
likely reflects an expansion before the freeze-out density is reached.
This picture resembles the dynamics of the EES model which has an expansion
phase before the bulk of the IMF's are emitted. This  expansion phase might
gradually happen faster for higher incident energies and most likely for 
heavier projectiles as indicated in the measured energy spectra.

The energy spectra of the IMF's turn out to be sensitive observables.
In p+Au collisions, the energy spectra are well described by the
empirically modified cascade-SMM calculations. However, for
$^4$He and $^{12}$C induced reactions the spectra exhibit stronger high-energy
tails which cannot be reproduced
by the calculations. This effect cannot be accounted for
by any variation of the residual masses. We attribute this observation
to the persistence of collective flow in the system at freeze-out.
Assuming a linear radial profile for the
flow velocity, its value at the surface is estimated to be around
$0.1c$ both for $^4$He and $^{12}$C induced reactions.
However, a detailed inspection of the variation of the kinetic energies of
the fragments reveals that the flow velocities seem to vary with the fragment
charge.
This is in contrast to model expectations which assume equal probability
for fragments of a given charge to be formed at any point in the
break-up volume.
This discrepancy
indicates that heavier fragments are formed more
in the interior of the system, possibly due to a density gradient.

This study of multifragmentation using a range of projectiles from protons
 to light nuclei provides new information
on several aspects of multifragmentation. It demonstrates
a transition from a pure ``thermal decay'' (for p+Au Collisions)
to a disintegration characterized by a collective flow (for heavier
projectiles). Nevertheless, the decay mechanism looks like a thermal
multifragmentation, as the partition of the system is governed by the
nuclear temperature. In all cases the IMF charge distributions are
well described by statistical multifragmentation models without flow.
This transition occurs at incident energies around 6 GeV using light
projectiles.  Our study puts into question the usual two-stage concept for
describing multifragmentation.  Furthermore, a realistic, non-uniform
density distribution might be needed to properly describe the phase-space
occupancy at freeze out.

The authors are thankful to Profs. A.~Hrynkiewicz, A.N.~Sissakian, 
A.I.~Mala\-khov, N.A.~Russakovich and S.T.~Belyaev for support. 
The research was supported in part by Grant
No 00-02-16608 from Russian Foundation for Basic Research, by Grant
No 2P03 12615 from the Polish State Committe for Scientific Research 
and Grant of the Polish Plenipotentiary in JINR,
by Grant
No 94-2249 from INTAS, by Contract No 06DA819 with Bundesministerium
f\"ur Forschung und Technologie, by Grant PST.CLG.976861 from NATO,
and by US National Science Foundation.

\newpage


\end{document}